\documentclass[iop, apjl, twocolumn]{emulateapj}
\slugcomment{Revised manuscript submitted to ApJ Letter}

\usepackage{amsmath}

\shorttitle{Electron Acceleration in Solar Flares}
\shortauthors{Petrosian \& Chen}

\begin{document}

\title{Derivation of Stochastic Acceleration Model Characteristics 
for Solar Flares From {\it RHESSI} Hard X-Ray Observations}
\author{Vah\'e Petrosian\altaffilmark{1} and Qingrong Chen}
\affil{Department of Physics, Kavli Institute for Particle Astrophysics and Cosmology, 
Stanford University, Stanford, CA 94305; vahep@stanford.edu, qrchen@gmail.com}
\altaffiltext{1}{Also Department of Applied Physics, Stanford University.}

\begin{abstract}
The model of stochastic acceleration of particles by turbulence has been
successful in explaining many observed features of solar flares.
Here we demonstrate a new method to obtain the accelerated electron spectrum
and important acceleration model parameters
from the high resolution hard X-ray observations provided by
the {\it Reuven Ramaty High Energy Solar Spectroscopic Imager} ({\it RHESSI}).
In our model, electrons accelerated at or very near the loop top 
produce thin target bremsstrahlung emission there and then
escape downward producing thick target emission at the loop footpoints.
Based on the electron flux spectral images obtained by the regularized inversion of
the {\it RHESSI} count visibilities, 
we derive  several important parameters for the acceleration model.
We apply this procedure to the 2003 November 03 solar flare, which
shows a loop top source up to 100--150 keV in hard X-ray with a relatively flat spectrum
in addition to two footpoint sources.
The results imply presence of strong scattering and a high density of turbulence energy
with a steep spectrum in the acceleration region.
\end{abstract}


\section{Introduction}
It is well established that the impulsive phase hard X-ray (HXR) emission of
solar flares is produced by bremsstrahlung  of nonthermal electrons spiraling
down the flare loop  while losing energy primarily via elastic Coulomb collisions
\citep{Brown71, Hudson72, Petrosian73}.
Thus, HXR observations provide the most direct information on the spectrum of
the radiating electrons and perhaps on the mechanism responsible for their acceleration.
The common practice to extract this information has been to use the {\it
parametric forward fitting} of HXR spectra to emission by an assumed spectrum, 
usually a power-law with breaks and cutoffs (or plus a thermal component),
of the radiating or accelerated electrons \citep[e.g.][]{Holman03}.
A more direct connection was established between the observations and the acceleration process
first by \citet{Hamilton92}, fitting to high spectral resolution
but narrow band observations \citep{Lin87},
and later by \citet{Park97},  fitting to broad band observations
\citep[e.g.][]{Marschhauser94, Dingus94}. 
This was done in the framework of stochastic acceleration (SA) by plasma waves or turbulence.

However, it is preferable to obtain the X-ray radiating electron spectrum
{\it nonparametrically} by some inversion techniques first attempted by \citet{Johns92}.
Recently, \citet{Piana03} and \citet{Kontar04}
applied regularized inversion techniques to obtain
the radiating electron flux spectra from the spatially integrated
photon spectra observed by {\it RHESSI} \citep{Lin02}.
This is an important advance but it gives the spectrum of the effective
{\it radiating electrons} summed over the whole flare loop,
but not the spectrum of the {\it accelerated electrons}.
This difference arises because high spatial resolution observations, first from
{\it Yohkoh} \citep{Masuda94, Petrosian02} and now from {\it RHESSI} \citep[e.g.][]{LiuW03},
have shown that, essentially for all flares,
in addition to the  emission from the loop footpoints (FPs) \citep[e.g.][]{Hoyng81},
there is substantial HXR emission from a region near the loop top (LT).
Thus, the total radiating electron spectrum is a complex
combination of the accelerated electrons at the LT and those present in the
FPs after having been modified by transport effects.

It is therefore clear that separate inversion of the LT and FP photon spectra
to electron spectra would provide more direct information on the acceleration mechanism.
More recently, \citet{Piana07} have applied the regularized inversion technique
to the {\it RHESSI} data in the Fourier domain \citep{Hurford02} to obtain electron flux spectral images.
The goal of this letter is to demonstrate that with
the resulting spatially resolved electron flux spectra at the LT and FPs 
one can begin to constrain the acceleration model parameters directly.

In the next section we present a brief review of the relation between the
derived electron flux images and the characteristics of the SA model and in \S3 we
apply this relation to a flare observed by {\it RHESSI}. A brief summary and
our conclusion are presented in \S4.

\section{Acceleration and Radiation}
The observations of distinct LT and FP HXR emissions, with little or no
emission from the legs of the loop, point to the LT as the acceleration site
and require enhanced scattering of electrons in the LT.  
\citet{Petrosian99} showed that the most likely scattering agent
is turbulence which can also accelerate particles stochastically.
In fact SA of the background thermal plasma has 
been the leading mechanism for acceleration of electrons
\citep[e.g.][]{Hamilton92,Miller96, Park97, Petrosian04, Grigis06, Bykov09}
and ions \citep[e.g.][]{Ramaty79, Mason86, Mazur95, LiuS04, LiuS06, Petrosian09},
and  is the most developed model in terms of comparing with observations. 

\subsection{Particle Kinetic Equation}

In this model one assumes that turbulence is produced at or near the LT region
(with background electron density $n_{\rm LT}$, volume $V$, and size $L$). 
In presence of a sufficiently high density of turbulence the scattering
can result in a mean scattering length or time ($\tau_{\rm scat}$)
smaller than $L$ or the crossing time ($\tau_{\rm cross}=L/v$),
leading to a nearly isotropic pitch angle distribution \citep{Petrosian04}.
The general Fokker-Planck equation for the density spectrum $N(E)$
of the accelerated electrons, averaged over the turbulent acceleration region,
simplifies to
\begin{eqnarray}
\frac{\partial N}{\partial t}=\frac{\partial^2}{\partial E^2}\left[D_{\rm EE}N\right]
&-&\frac{\partial}{\partial E}\left[\left(A(E)-\dot{E}_{\rm L}(E)\right)N\right]\nonumber \\ 
&-&\frac{N}{T_{\rm esc}(E)}+\dot{Q}(E),
\label{FPEq}
\end{eqnarray}
where $D_{\rm EE}(E)$ and $A(E)$ are the diffusion rate and direct acceleration rate 
by turbulence\footnote{For stochastic acceleration, 
$A(E)=D_{\rm EE}\zeta(E)/E+ {\rm d} D_{\rm EE}/{\rm d}E$,
where $\zeta(E)=(2-\gamma^{-2})/(1+\gamma^{-1})$,
$\gamma=1+E/m_{\rm e}c^2=1/\sqrt{1-\beta^2}$ is the Lorentz factor,
and $v=c\beta$ is the electron velocity.}, respectively,
$\dot{E}_{\rm L}$ is the electron energy loss rate,
and $\dot{Q}(E)$ and $N(E)/T_{\rm esc}(E)$ describe the rate of injection of
(thermal) particles and escape of the accelerated particles from the acceleration region.
For electrons of energies below $\sim$1 MeV, which are of interest here, 
Coulomb collisions\footnote{At higher energies, synchrotron loss 
must be included in $\dot{E}_{\rm L}$.}
dominate the energy loss rate, 
\begin{equation}
\dot{E}_{\rm L} = \dot{E}_{\rm Coul}=4\pi r_0^2 m_{\rm e} c^3 n_{\rm LT}\ln\Lambda/\beta,
\end{equation}
where $\ln\Lambda$ is the Coulomb logarithm taken to be 20 for solar flare conditions.
Following \citet{Petrosian04}, we approximate the escape time as
$T_{\rm esc}(E)\simeq\tau_{\rm cross} ( 1+\tau_{\rm cross}/\tau_{\rm
scat})$, which smoothly connects the two limiting cases of
$\tau_{\rm cross}/\tau_{\rm scat}\gg 1$ and $\ll 1$.
The mean scattering time is related to the pitch angle diffusion rates \citep{DP94, Pryadko97} 
due to both Coulomb collisions ($D^{\rm Coul}_{\rm \mu\mu}$ and $\tau^{\rm turb}_{\rm scat}$) 
and turbulence ($D^{\rm turb}_{\rm \mu\mu}$ and $\tau^{\rm Coul}_{\rm scat}$) as
\begin{equation}
\tau_{\rm scat}(E)=\frac{1}{8}\int_{-1}^1\frac{(1-\mu^2)^2}
{D_{\mu\mu}^{\rm Coul}(\mu,E)+ D_{\mu\mu}^{\rm turb}(\mu,E)} {\rm d}\mu.
\label{tausc0}
\end{equation}
Similarly we can define the scattering times 
$\tau^{\rm Coul}_{\rm scat}$ and $\tau^{\rm turb}_{\rm scat}$ for each process alone. 
For Coulomb collisions,
$D_{\mu\mu}^{\rm Coul}=\frac{2(1-\mu^2)}{\gamma+1}\frac{\dot{E}_{\rm Coul}}{{E}}$.
For turbulence, $D^{\rm turb}_{\mu\mu}$, like $D_{\rm EE}$, 
depends on the spectrum of turbulence and on
the background plasma density, composition, temperature, and magnetic field
\citep[see][]{Schlickeiser89, DP94, Pryadko97, Pryadko98, Pryadko99, Petrosian04}.
Since these coefficients determine the spectrum of the accelerated electrons,
one can then constrain some aspects of the acceleration mechanism if
an accurate spectrum of the electrons can be derived from  observations.

\subsection{LT and FP Spectra}

The accelerated electrons in the (LT) acceleration region with
a flux spectrum $F_{\rm LT}(E)=vN(E)$ produce {\it thin target}
bremsstrahlung emissivity (photons s$^{-1}$ keV$^{-1}$)
\begin{equation}
J_{\rm LT}(\epsilon)=n_{\rm LT}V \int_\epsilon^\infty F_{\rm
LT}(E)\sigma(\epsilon,E) {\rm d}E,
\label{LTbrem}
\end{equation}
where $\sigma(\epsilon,E)$ is the angle-averaged bremsstrahlung cross section \citep{KM59}.
The escaping electrons with flux $F_{\rm 0}(E)= N(E)L/T_{\rm esc}$
produce {\it thick target} bremsstrahlung emissivity (coming mostly from the FPs)
\citep[see][]{Petrosian73, Park97},
\begin{equation}
J_{\rm FP}(\epsilon)=n V \int_\epsilon^\infty 
F_{\rm FP}(E)\sigma(\epsilon,E) {\rm d}E,
\label{FPbrem}
\end{equation}
where $n$ is the density and 
$F_{\rm FP}$ is the effective radiating electron flux spectrum at the FPs,
\begin{equation}
F_{\rm FP}(E)=vN_{\rm FP}=\frac{v(E)}{\dot{E}_{\rm L}(n)}
\int_E^\infty\frac{N(E')}{T_{\rm esc}(E')}{\rm d}E'.
\label{effSpec}
\end{equation}
Since $\dot{E}_{\rm L}\propto n$,  the FP
photon spectrum is independent of density. In what follows we evaluate 
equations (\ref{FPbrem}) and (\ref{effSpec}) using the LT density $n_{\rm LT}$.

\subsection{Acceleration Model Parameters}

Regularized inversion of {\it RHESSI} count visibilities
gives the electron visibilities \citep{Piana07}, 
which can then be used to construct images of electron flux (multiplied by column density).
From these images, we extract the spatially resolved spectra, 
$F_{\rm LT}(E)$ at the LT and $F_{\rm FP}(E)$ at the FPs.
Thus we can obtain the accelerated electron spectrum $N(E)$ at the thin target LT.
Also from differentiation of equation (\ref{effSpec}) we derive the escape time as 
$T_{\rm esc}=-N(E)/\frac{\rm d}{{\rm d}E}(F_{\rm FP}\dot{E}_{\rm L}/v)$,
and by converting the denominator to a logarithm derivative we get
\begin{eqnarray}
T_{\rm esc}(E)=\frac{\tau_{\rm L}(E)(F_{\rm LT}/F_{\rm FP})}
{\delta_{\rm FP}(E)+2/(\gamma+\gamma^2)}\equiv \tau_{\rm L}(E)\xi(E),
\label{Tesc}
\end{eqnarray}
where the FP index $\delta_{\rm FP}(E)=-\frac{{\rm d}\ln F_{\rm FP}}{{\rm d}\ln E}$,
$1/(\gamma+\gamma^2)=-\frac{{\rm d}\ln v(E)}{{\rm d}\ln E}$, and 
$\tau_{\rm L}(E)=E/\dot{E}_{\rm L}$ is the Coulomb loss time at the LT.
The function $\xi(E)$ is an observable quantity representing the ratio $T_{\rm esc}/\tau_{\rm L}$. 
In the above derivation, we have used the relativistic form of electron velocity $v(E)$.

Given $T_{\rm esc}(E)$, from its relation to $\tau_{\rm cross}$ and $\tau_{\rm scat}$,
we obtain the mean scattering time as
$\tau_{\rm scat}\simeq \tau_{\rm cross}^2/(T_{\rm esc}-\tau_{\rm cross})$,
which is valid for $T_{\rm esc}>\tau_{\rm cross}$.
Disentanglement of $\tau^{\rm turb}_{\rm scat}$ from $\tau_{\rm scat}$ 
is complicated (eq. [\ref{tausc0}])
at energies when turbulence and Coulomb collisions contribute equally to $\tau_{\rm scat}$.
However, if turbulence dominates the pitch angle diffusion, 
then to the first order we can write
$\tau^{\rm turb}_{\rm scat} \simeq\tau_{\rm scat}(1+\tau_{\rm scat}/\tau^{\rm Coul}_{\rm scat})$,
and obtain some average value of $D^{\rm turb}_{\mu\mu}$.
Furthermore, given $N(E)$ we can in principle determine 
the other Fokker-Planck coefficients, namely $A(E)$ and $D_{\rm EE}$ (see eq. [\ref{FPEq}]).
Therefore we can reach a consistent picture of the acceleration process
due to turbulence and begin to make inroads into the spectrum and the nature of
turbulence itself.

\begin{figure*}[t]
\epsscale{1.}\plotone{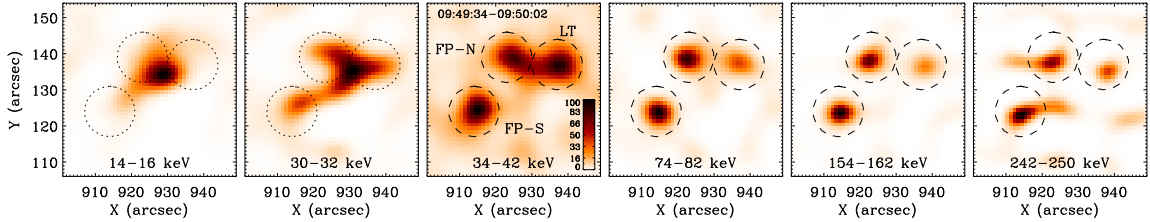}
\caption{Electron flux spectral images (with 8 keV bin width above 34 keV and 2 keV bin width at lower energies)
up to 250 keV in the 2003 November 03 flare during the nonthermal peak as reconstructed 
from two sets of the regularized electron visibilities by the MEM\_NJIT algorithm \citep{Schmahl07}.
The images show one LT and two FP sources above 34 keV and a loop structure at lower energies. 
Three circles are used to extract the LT and FP electron flux spectra
above 34 keV (see Figure \ref{Nov03_Spectra}).
}
\label{Nov03_Images}
\end{figure*}

\begin{figure*}[bhpt]
\epsscale{0.52}\plotone{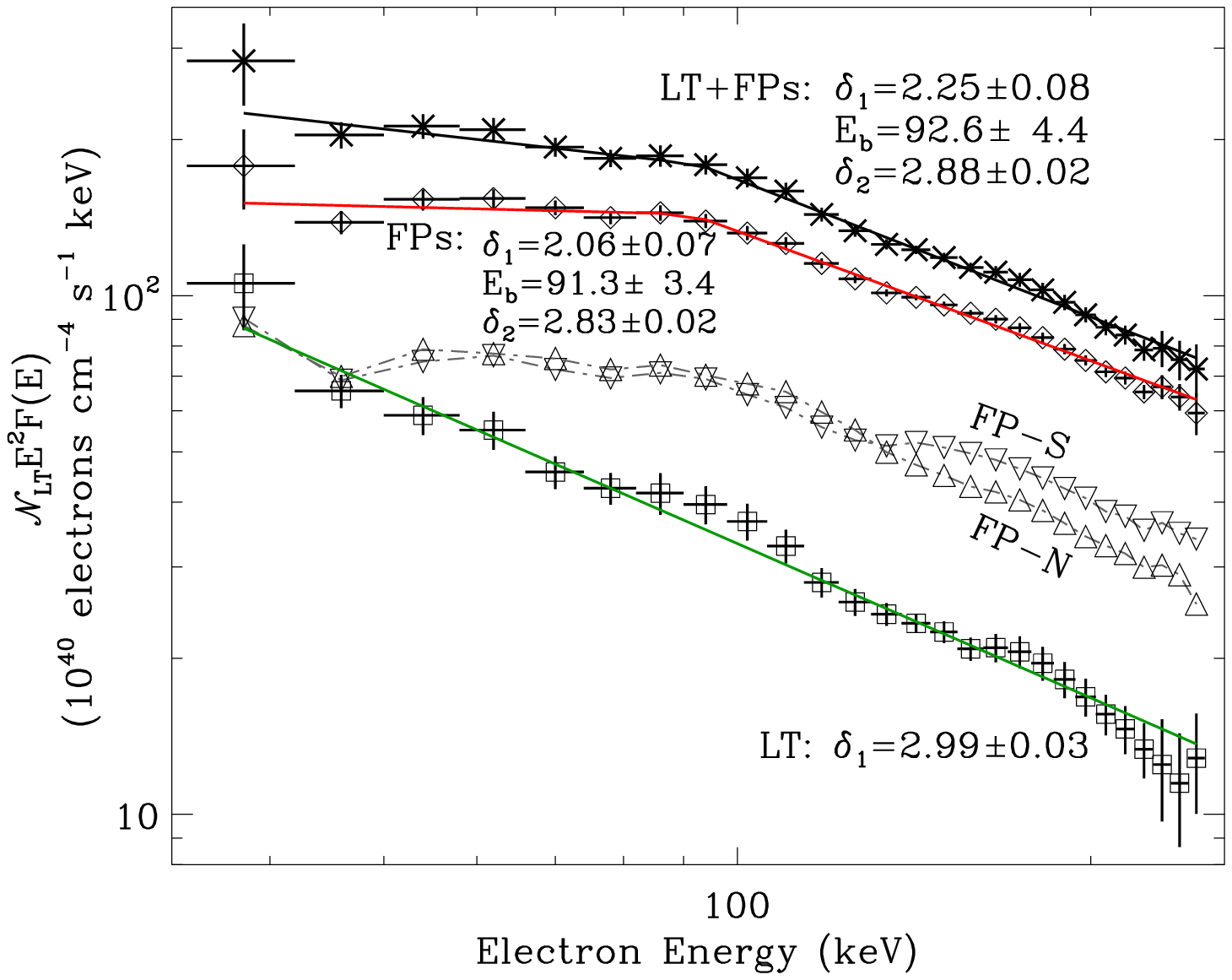}
\epsscale{0.52}\plotone{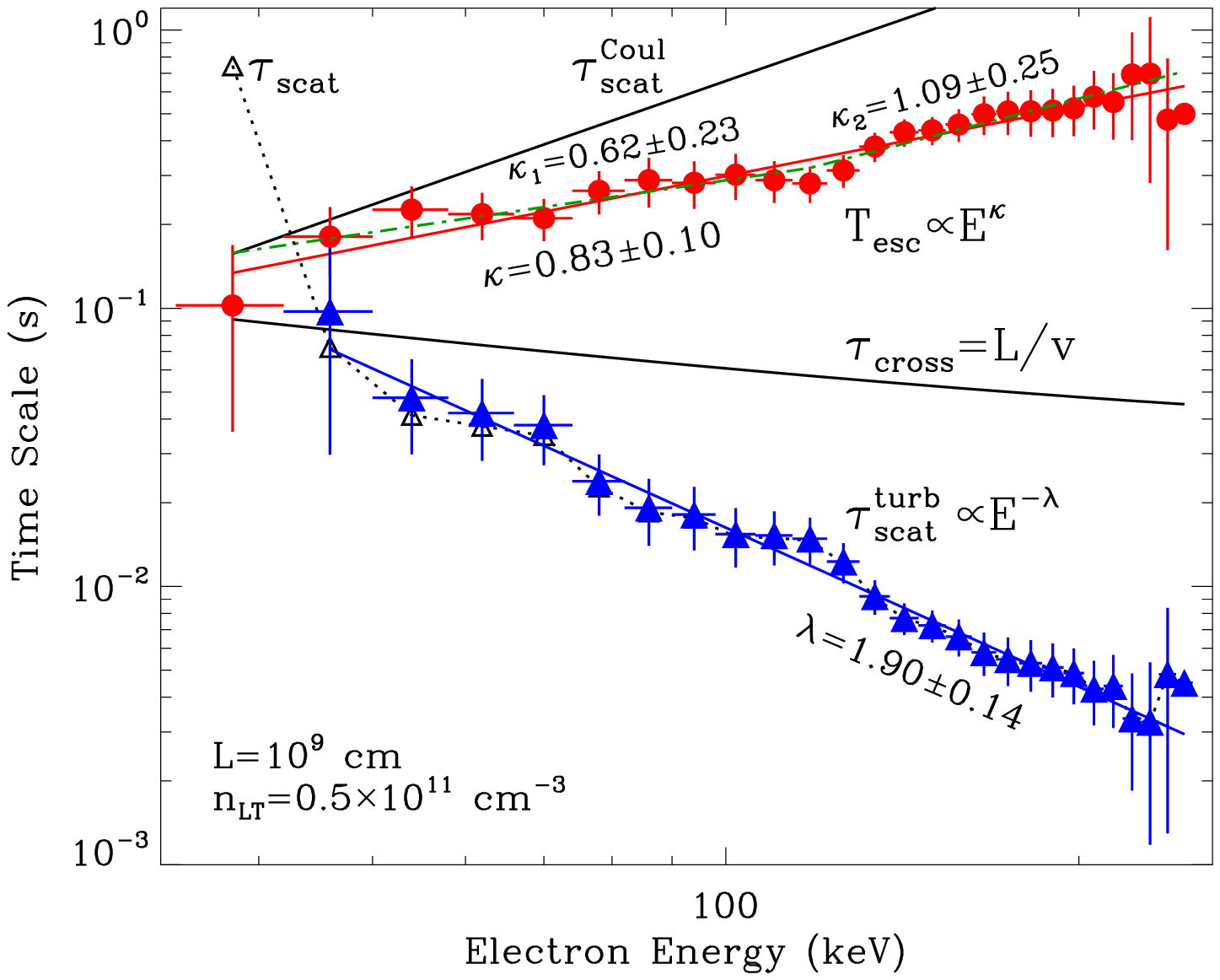}
\caption{
{\it Top}:
Electron power spectra ${\cal N}_{\rm LT}E^2F(E)$ for the LT ({\it square}),
the two FPs summed ({\it diamond}), and all three sources (LT + FPs, {\it cross}) 
in the 2003 November 03 flare. The LT spectrum can be fitted by a power-law, 
and the summed FP and total spectra by a broken power-law.
Also note that the southern FP spectrum ({\it downward triangular}) is flatter than 
the northern FP spectrum ({\it upward triangular}), 
mostly above $\sim$90 keV by $\sim$0.3 powers of energy,
consistent with their asymmetric locations with respect to the LT.
{\it Bottom}:
Escape time ({\it filled circle}) 
and turbulence scattering time ({\it filled triangular}) in the (LT) acceleration region. 
The escape time can be well fitted by either a power-law 
or a broken power-law ({\it dash dot}) increasing with energy,
and the turbulence scattering time by a power-law rapidly decreasing with energy.
Also shown are the crossing, Coulomb scattering, and the mean scattering 
({\it open triangular}) times.
The reduced chi-squares for all the fittings are below or around 1.
}
\label{Nov03_Spectra}
\end{figure*}

\section{Application: The 2003 November 03 Flare}

As a first demonstration, we apply our new procedure to 
the 2003 November 03 solar flare (X3.9 class) during the nonthermal peak, 
in which we find a hard LT source (extending above 100 keV in HXR) 
distinct from the thermal loop in addition to two FP sources\footnote{
Q. Chen \& V. Petrosian (2010a, in preparation) 
present HXR observations of this flare and argue that
the high energy LT source should not be an artifact of the pulse pileup effect.}.
In Figure \ref{Nov03_Images}
we show the electron flux images up to 250 keV,
which also show a loop at low energies and one LT and two FPs at higher energies.
In Figure \ref{Nov03_Spectra} {\it top} panel we show the electron spectra
${\cal N}_{\rm LT}E^2F(E)$, where ${\cal N}_{\rm LT}=n_{\rm LT}L$ is the LT column density.
The LT flux spectrum can be fitted by a power-law with an index $\delta_{\rm LT}=3.0$.
The summed FP flux spectrum can be better fitted by a broken power-law with the indexes $\delta_1= 2.1$
and $\delta_2= 2.8$ below and above the break energy $E_{\rm b}= 91 \pm 3$ keV.
It is clear that the total radiating electron spectrum 
differs significantly from the (LT) accelerated electron spectrum.

Given the above LT and FP electron flux spectra we derive the
energy dependence of the escape time (eq. [\ref{Tesc}]).
The LT density can be  estimated as $n_{\rm LT}\simeq \sqrt{{\rm EM}/L^3}\simeq
0.5\times10^{11}\ {\rm cm}^{-3}$, where
the LT size $L \simeq 10^9$ cm is  obtained from the LT angular size,
and the emission measure ${\rm EM}\simeq 0.2\times 10^{49}$ cm$^{-3}$ is
obtained from spectral fitting of the LT thermal emission.
As in Figure \ref{Nov03_Spectra} {\it bottom} panel,
the escape time increases slowly with energy 
and can be fitted by either a power-law,
\begin{equation}
T_{\rm esc}(E) = 0.3\ {\rm s} \left(\frac{E}{100\ {\rm
keV}}\right)^{\kappa},\,\,\,
\kappa=0.83\pm0.10,
\end{equation}
or a broken power-law with a break at $E_{\rm b}=118\pm37$ keV,
and the indexes $\kappa_1=0.62\pm0.23$ and $\kappa_2=1.09\pm0.25$.
The fact that the escape time should be longer than the crossing time
yields an upper limit on ${\cal N}_{\rm LT}$,
which is satisfied by the above LT density and size.

We then calculate the mean scattering time in the LT region. 
Except at the lowest energy, the Coulomb contribution is small 
so that the scattering time thus calculated can be attributed to turbulence. 
The scattering time due to turbulence (see \S2.3) can be fitted 
by a power-law above $\sim$40 keV,
\begin{eqnarray}
\tau^{\rm turb}_{\rm scat}=0.016\ {\rm s}
\left(\frac{E}{100\ {\rm keV}}\right)^{-\lambda},
\,\,\, \lambda=1.90\pm 0.14.\,\,
\end{eqnarray}

\section{Summary and Discussion}

In this paper we describe a new method to directly obtain the model parameters
for stochastic acceleration of particles by turbulence in solar flares from
regularized inversion of the high resolution {\it RHESSI} HXR data \citep{Piana07}.
We have argued that particle acceleration takes place at or near the LT region.
The accelerated electrons produce thin target emission at the LT
and then escape downward to the dense FP region
undergoing Coulomb collisions and producing thick target emission.
In this model the LT and FP electron spectra are connected by
the escape process from the LT region (eq. [\ref{effSpec}]),
thus allowing us to determine the energy dependence of the escape time.
Our method has the advantage that one can now constrain the model
parameters uniquely rather than just satisfying the consistency between the
model and the data as commonly done by forward fitting routines.
This method can be applied to flares with simultaneous HXR emission from the LT and FP sources.

We have applied our method to the 2003 November 03 flare, in which
we can obtain the electron flux images for both the LT and FPs up to 250 keV.
The LT accelerated electron flux spectrum can be fitted by a power-law
and the effective radiating flux spectrum at the FPs is better fitted
by a broken power-law.
From these spectra we derive the energy variation of the escape time and the scattering time.
As seen in Figure \ref{Nov03_Spectra},
the turbulence scattering time is relatively short and decreases with energy.
A  short scattering time may arise from
a high energy density of turbulence (${\cal E}_{\rm turb}$),
with the exact relationship depending also on the magnetic field ($B$),
and the spectral index ($q$) and minimum wave number ($k_{\rm min}$) of turbulence.
A  high level of turbulence also implies efficient acceleration
which generally means a flat spectrum
for the accelerated electrons, which is the case for the current flare.
The energy dependences of $\tau^{\rm turb}_{\rm scat}$ and $D_{\rm EE}$
are also a function of these characteristics of turbulence;
at high energies they are determined primarily by the spectral index
of turbulence \citep[see][]{DP94, Pryadko97, Pryadko98, Pryadko99, LiuS06}.

For the usually assumed Kolmogorov ($q=5/3$)
or Iroshnikov-Kraichnan ($q=3/2$) turbulence spectra,
one expects the scattering time to increase with energy as $E^{2-q}$,
which translates into an escape time varying roughly as $T_{\rm esc}\propto 1/\sqrt{E}$
at high (but non-relativistic) energies.
The energy dependences of $T_{\rm esc}$ and $\tau^{\rm turb}_{\rm scat}$ obtained here
require a steeper turbulence spectrum ($q > 3$) at high wave numbers.
Such a steep spectrum can be present beyond the inertial range where damping is
important \citep[e.g.][]{Jiang09}.
The electron energies and the wave-particle resonance condition
determine the wave vector of the accelerating plasma waves.
This relation depends primarily on the plasma parameter
$\alpha\propto \sqrt{n}/B$ \citep[e.g.][]{Petrosian04}.
Thus, given the magnetic field and plasma density we can determine the wave vectors
for transition from the inertial to  the damping  ranges of turbulence.

It should, however, be emphasized that the results obtained here may
not be representative of typical flares.
More commonly flares have much softer LT emission,
which would give an escape time decreasing (and scattering time increasing) with
energy, consistent with a low level and  a flat spectrum of turbulence.

The exact relation between the derived quantities 
($N(E)$, $T_{\rm esc}$, and $\tau^{\rm turb}_{\rm scat}$)
and the turbulence characteristics (${\cal E}_{\rm turb}, B, q$, etc.)
is complicated and depends on the angle of propagation of the plasma waves
with respect to magnetic field and other plasma conditions.
In future, we will apply these procedures to more flares 
(Q. Chen \& V. Petrosian, 2010b, in preparation)
and deal with these relations explicitly.

\acknowledgments
We thank the referee for helpful comments.
We thank Anna Maria Massone and Gordon Hurford for providing 
the visibility inversion code and valuable discussions about data analysis, 
and Siming Liu and Wei Liu for various discussions.
{\it RHESSI} is a NASA small explorer mission. 
This work is supported by NSF grant ATM0648750 and NASA grant NNX10AC06G.


\end{document}